\documentclass[a4,11pt]{article}
\usepackage{amsmath,amssymb,amsthm}
\usepackage{graphicx}
\usepackage{amsmath,amsthm,amstext,amscd,amssymb,euscript,mathrsfs}
\usepackage{epsfig}
\def\1{{\mathchoice {\rm 1\mskip-4mu l} {\rm 1\mskip-4mu l}
{\rm 1\mskip-4.5mu l} {\rm 1\mskip-5mu l}}}

\newtheorem{THM}{Theorem}[section]
\newtheorem{PRP}[THM]{Proposition}
\newtheorem{LEM}[THM]{Lemma}

\newtheorem{EXA}{Example}

\newtheorem{REM}[THM]{Remark}

\newtheorem{COR}[THM]{Corollary}
\newtheorem{DEF}[THM]{Definition}

\newcommand{\R}{\mathbb{R}}

\newcommand{\E}{\mathbb{E}}
\newcommand{\Z}{\mathbb{Z}}
\newcommand{\N}{\mathbb{N}}
\newcommand{\C}{{\bf C}}

\newcommand{\blank}[1]{}

\title{Propagation of Gibbsianness for infinite-dimensional diffusions with space-time interaction}
\author{S. R\oe lly$^{\textup{{\tiny(a)}}}$ and W.M. Ruszel$^{\textup{{\tiny(b)}}}$ \\
{\small $^{\textup{(a)}}$
Universit\"at Potsdam,}\\
{\small Institut f\"ur Mathematik}\\
{\small Am Neuen Palais 10
, D-14469 Potsdam, Germany}\\
{\small $^{\textup{(b)}}$ Technical University Delft}\\
{\small Delft Institute of Applied Sciences}\\
{\small Mekelweg 4,
2628 CD Delft, The Netherlands}}
\begin{document}
\maketitle

\begin{quote}
Abstract: We consider infinite-dimensional diffusions where the interaction between the coordinates has a finite extent both in space and time. 
In particular, it is not supposed to be smooth or Markov.
The initial state of the system is Gibbs, given by a strong summable interaction.
If the strongness of this initial interaction is lower than a suitable 
level, and if the dynamical interaction is bounded from above in a right way,
we prove that 
the law of the diffusion  at any time
$t$ is a Gibbs measure with absolutely summable interaction.
The main tool is a cluster expansion in space uniformly in time of the Girsanov factor coming from the
dynamics and exponential ergodicity of the free dynamics to an equilibrium product measure.
\end{quote}

\newpage

\section{Introduction}

In this paper we study propagation of Gibbsianness for a class of 
infinite-dimensional diffusions with general space-time interaction. 
The diffusion $X=(X_i(t))_{t \geq 0, i \in \Z^d}$ solves the
 Stochastic Differential Equation \eqref{SDE}
where the dynamical interaction splits into a suitable self-interaction and a  bounded (possibly) non-regular space interaction with time memory.\\

Diffusions with memory, such as stochastic delay equations are indeed very useful for stochastic
modeling e.g. in biomathematics, mathematical finance or physics, where
delays in the dynamics can represent memory, inertia in financial systems or time-delayed response of physical systems
(see e.g.  \cite{DvGVW95}, \cite{HV05}, \cite{AHMP07}, \cite{WSB04} or \cite{TP01}).\\

Recall that simple transformations of Gibbs measures may not preserve the Gibbsianness property. 
The phenomenon was 
identified by van Enter, Fernand\'ez and Sokal in \cite{VEFS93} and, since then, an extensive effort has been made to 
find various situations where such pathologies may arise.
An example
of a transformation which could yield non-Gibbs measures is time-evolution. More precisely, consider a system of interacting
particles (or spins) living on a certain space $S$ and distributed at time $t=0$ according to some Gibbs measure $\nu$. It may happen that, although the
system converges, as time
goes to infinity, towards another Gibbs measure $\mu$, under certain conditions on $\nu$ and $\mu$, there exists a period of time where
the time-evolved measure  is not Gibbs any more, since an associated absolutely summable interaction does not exist. Such unexpected behavior was pointed out in
the following cases: for discrete state space $S$ and spin-flip dynamics in \cite{VEFHR02, VEEIK12}, in the mean-field set-up, see \cite{KLN07,EK10},
for Markovian diffusions on circles, called planar rotors, in \cite{VER08, VER09} and for continuous unbounded spins following independent
Ornstein-Uhlenbeck dynamics, see \cite{KR06}. Note that dynamical Gibbs-non-Gibbs transitions have also been investigated from a large-deviation point
of view in \cite{VEFHR10}.

Here, on the contrary, we are interested in the {\em conservation of Gibbsianness} regime for particles living in continuous state spaces. We search for 
conditions which assure that the time-evolved measure of a
 system of interacting particles starting from a Gibbs distribution stays Gibbsian during its {\em whole time-evolution}. \\
It turns out that for short-time evolutions, conservation of Gibbsianness is robust, as it was proved in \cite{DR05} for Markovian
$\mathbb{R}^{\mathbb{Z}^d}$-valued diffusions and in \cite{RRR10} for a particular  class of non-Markovian $\mathbb{R}^{\mathbb{Z}^d}$-valued diffusions.
Earlier propagation of Gibbsianness results during the whole time-evolution could already be obtained in \cite{DR05} in  the following particular case: the 
$\mathbb{R}^{\mathbb{Z}^d}$-valued diffusion is prescribed through a {\em Markov} interaction function $b$, itself defined as the gradient of a Hamiltonian. \\

Consider a diffusion $X=(X_i(t))_{t \geq 0, i \in \Z^d}$ where the dynamical interaction term consists of an ultracontractive self-interaction $U$ (which 
will constrain the free system to converge fast towards a reference product measure)  
and a bounded non-regular and non-Markov space-time interaction $b$ regulated by a multiplicative scalar factor $\beta$. \\
We prove that, for any initial Gibbs measure with inverse temperature $\beta_0$ bounded above  ($\beta_0 < \bar\beta_0 $), and for a dynamical interaction 
under a certain intensity ($\beta < \bar\beta $), 
the law of the diffusion at any time $t$ is a Gibbs measure on $\mathbb{R}^{\mathbb{Z}^d}$, described by an absolutely summable interaction. In
that sense,
Gibbsianness propagates for a very large class of $\mathbb{R}^{\mathbb{Z}^d}$-valued diffusion dynamics which include time-delayed terms.  \\
As a corollary, 
our
method leads to a constructive existence result for a class of infinite-dimensional SDE with small (possibly) non-Markovian drift. Finally in section
\ref{secAppl} we state the corresponding  propagation of Gibbsianness result for a system of planar rotors.

There are in the present paper  two main differences and improvements with respect to the paper \cite{DR05}. 
First, the Girsanov  density of the approximate finite-dimensional diffusions
contains stochastic integrals, which
cannot be turned into ordinary (bounded) integrals as was done for gradient diffusions. In particular the local interaction functionals $\Psi$
introduced in (\ref{Psi}) 
are highly unbounded, even not everywhere defined and their  control should be done via (exponential) moments.
Secondly, since the interaction $b$ between the coordinates contains a time component, one cannot make use any more 
of the decoupling method as in \cite{DR05}, which was a simple way to compare the infinite-dimensional dynamics with another one much simpler. 
To bypass these difficulties, the main tool is
a cluster expansion in space - uniform in time - of Girsanov factors coming from the dynamics
.\\

 The rest of the paper is divided into the following sections.
\ref{secFrame}. Framework and main result.
\ref{secProof}. Proof of the main theorem with, in particular, the cluster expansion and the estimates of the cluster weights. 
In section \ref{secAppl} we come back to examples and applications.

\section{Framework and main result}\label{secFrame}

In this section we define the necessary framework for our study and state our main result.

\subsection{Interaction and Gibbs measures}
The main mathematical concept considered in this paper is that of Gibbs measure on the configuration space  $\mathbb{R}^{\mathbb{Z}^d}$. 
It is based on a so-called interaction function, whose we now recall the definition.
\begin{DEF}\label{interac}
An {\bf interaction} $\phi$ on $\mathbb{R}^{\Z^d}$ is a collection of functions $\phi_{\Lambda}$ from
$\mathbb{R}^{\Z^d}$ to $\mathbb{R}$, where $\Lambda$ is any finite subset of $\Z^d$,  satisfying the
following properties. 
\begin{enumerate}
\item $\phi$ is $\mathcal{F}_{\Lambda}$-measurable, where $\mathcal{F}_{\Lambda}$ denotes the sigma-field generated by the canonical projections
on $\mathbb{R}^{\Lambda}$
\item $\phi$ is absolutely summable, which means that, for all $i \in \Z^d$, $\quad \sum_{\Lambda \ni i} ||\phi_{\Lambda}||_{\infty} < + \infty$.
\end{enumerate}
\end{DEF}
We also recall some other summability assumptions which can be satisfied by an interaction.
\begin{enumerate}
\item[(A1)] (strong summability) $\sup_{i \in \mathbb{Z}^d} \sum_{\Lambda \ni i} (|\Lambda| -1) ||\phi_{\Lambda}||_{\infty} < + \infty$, where $|\Lambda|$
denotes the cardinality of $\Lambda$ .
\item[(A2)] (finite-body interaction) $\phi_{\Lambda} \equiv 0$ as soon as $|\Lambda|$ is large enough.
\item[(A3)] (finite-range interaction) $\phi_{\Lambda} \equiv 0$ as soon as the diameter of $\Lambda$ is large enough.
\end{enumerate}
Remark that (A3)$\Rightarrow$ (A2)$\Rightarrow$ (A1).\\
 
Given an interaction $\phi$ we define the associated Hamiltonian function $h=(h_{\Lambda})_{\Lambda \subset \Z^d}$by
\begin{equation}
h_{\Lambda}: \R^\Lambda \times \R^{\Lambda^c} \rightarrow \R, \quad  h_{\Lambda}(x_{\Lambda},z_{\Lambda^c})= 
 \underset{\Lambda^{\prime}: \Lambda^{\prime} \cap \Lambda \neq \emptyset}\sum \phi_{\Lambda^{\prime}}(x_{\Lambda}z_{\Lambda^c}),
\end{equation}
where $z$ is called the  boundary condition. We write as usual $x_{\Lambda}z_{\Lambda^c}$ as shorthand for the concatenation of the
configuration $x$ restricted to $\Lambda$ and the 
configuration $z$ restricted to $\Lambda^c $.

The \textbf{finite-volume Gibbs measure} with interaction $\phi$  at inverse temperature $\beta_0$ with boundary condition $z$ w.r.t. an a-priori measure $m$ on
$\R$ is the
probability measure given by
\begin{equation}
\nu_{\Lambda, z}(dx_{\Lambda}) = \frac{1}{Z_{\Lambda}^z} \exp (-\beta_0 h_{\Lambda}(x_{\Lambda},z_{\Lambda^c})) \, m^{\otimes
\Lambda}(dx_{\Lambda})\label{gibbs}
\end{equation}
where $Z_{\Lambda}^z $ is the renormalizing factor.
If the measure $m$ is finite, the scalar $Z^z_{\Lambda}$,   also called partition function, is finite too. \\
As usual the finite-volume measure with \textbf{free boundary conditions} is defined by
\begin{equation}
\nu_{\Lambda}(dx_{\Lambda}) =  \frac{1}{Z_{\Lambda}} \exp \bigl(- \beta_0 \sum_{A \subset \Lambda} \phi_A(x_{\Lambda}) \bigr) \, m^{\otimes
\Lambda}(dx_{\Lambda}).
\end{equation}
We can now define the concept of (infinite-volume) Gibbs measure.
\begin{DEF}
The measure $\nu$ is a Gibbs measure with interaction $\phi$ at inverse temperature $\beta_0$  if for all finite $\Lambda \subset \Z^d$
and smooth  $\mathcal{F}_{\Lambda}$-measurable test functions $f$, the so-called DLR equations are satisfied
\begin{equation}
\int f(x_{\Lambda}) \, \nu(dx) = \int \int f(x_{\Lambda}) \, \nu_{\Lambda, z} (dx_{\Lambda})\,  \nu(dz), \label{DLR}
\end{equation}
which means that $\nu_{\Lambda, z} $ is a regular version of the conditional probability $\nu(dx_{\Lambda}|x_{\Lambda^c}=z_{\Lambda^c})$.
One denotes by  $\mathcal{G}_{\beta_0}(\phi)$ the set of such Gibbs measures.
\end{DEF}

\subsection{Infinite-volume dynamics}
On the path space $\Omega=C(\R_+,\mathbb{R})^{\Z^d}$, endowed by the canonical
sigma-field $\mathcal{F}$,  we consider the infinite-dimensional diffusion defined as solution of the Stochastic Differential Equation:
\begin{equation} 
 \begin{cases}
 & dX_i(t) = dB_i(t) + \big( -\frac{1}{2}U^{\prime}(X_i(t)) + \beta \, b_i(t,X) \big)\, dt, \,  i \in \mathbb{Z}^d \\
 & X(0) \sim \nu , \label{SDE}
 \end{cases}
\end{equation}
where $(B_i)_{i \in
\mathbb{Z}^d}$ is a sequence of real-valued independent Brownian motions, 
 $U$ is a  self-potential function, and 
the drift term of the $i^{th}$ coordinate  at time $t$, $b_i(t,\cdot)$, may possibly
depend on the values of the other coordinates of the process on the whole time interval $[0,t]$. Thus the process $X$ could be non-Markov. \\
We denote by $Q^\nu$ the law of the solution
of the SDE \eqref{SDE} (resp. $Q^{x}$ if the initial condition is deterministic, i.e. $\nu=\delta_x$). \\

We now state the precise assumptions satisfied by the drift term.
\begin{itemize}
\item[(B1)]  The self-potential $U: \R\rightarrow \R$ is smooth and \textit{ultracontractive}, in such a way that 
 the one-dimensional free dynamics
\begin{equation} \label{freedyn1_dim}
dx(t) = dB(t) - \frac{1}{2}U^{\prime}(x(t))dt
\end{equation} 
generates a semi-group which  maps $L^2(m)$ into $L^\infty(m)$, where $m$ 
is its unique stationary probability measure: $m(dx)=\frac{1}{Z}e^{-U(x)}dx$.
\item[(B2)] The space-time interaction is the product of a scalar intensity parameter $\beta$ 
with a functional $b=(b_i)_i$ on $\Omega$ which is adapted and local
in space and time:
There exists a finite neighborhood $\mathcal{N} \subset \Z^d$ around 0  and a finite memory-time $t_0>0$ such that 
 $$
\forall i \in \Z^d, \, \forall \omega \in \Omega, \quad b_{i}(t,\omega)= b_i(t,(\omega_{i +
\mathcal{N}}(s):t-t_0\leq s\leq t)).
$$
\item[(B3)] The drift functional  $b$ is bounded, i.e. 
$$
\exists \bar b > 0 \textrm{ such that } \underset{i \in \Z^d}\sup\, \underset{\omega \in
\Omega}\sup  \,  \underset{t \geq 0}\sup \, |b_i(t,\omega) | \leq \bar b.
$$
\end{itemize}

The following theorem is the main  result of our paper.
\begin{THM} \label{mainTh}
Consider $Q^{\nu}$, the law of the infinite-dimensional SDE \eqref{SDE} 
with a drift satisfying assumptions {\em (B1)-(B3)} and suppose that the initial distribution 
 $\nu$ is a Gibbs measure in $\mathcal{G}_{\beta_0 }(\phi)$ where $\phi$ satisfies the strong summability assumption {\em (A1)}.
There exists a bound $\bar \beta_0>0$ for the initial inverse temperature and a bound  $\bar \beta > 0$ 
for the intensity of the space-time interaction such
that,
if  $0 \leq \beta \leq \bar\beta$ and $ 0 \leq \beta_0
\leq \bar\beta_0$, for all $t \geq 0$ the time-evolved
measure $Q^{\nu}\circ X(t)^{-1}$ is a Gibbs measure  w.r.t. some interaction $\phi^t$, which is then absolutely summable.
\end{THM}

\begin{COR}\label{mainCor}
The above Theorem \ref{mainTh} provides a constructive way to obtain a solution of the SDE \eqref{SDE} at any time $t$ for small $\beta$ as limit (in terms
of cluster expansions) of finite-dimensional approximations, whose existence (and uniqueness) is ensured by the assumption {\em (B3)}.
\end{COR}

\section{Proof}\label{secProof}

The dynamics we deal with are obtained  by perturbing through the interaction $\beta \, b$ a system of independent evolving components. 
The law on  $\Omega$ of the non-interacting system called also infinite-dimensional {\it free system}, corresponding to $\beta=0$ and the deterministic
initial value $x \in \R^{\Z^d}$,  is denoted by $P^{x}$ and is the product law 
$$
P^x=\otimes_{i \in \Z^d} P^{x_i}_i
$$
where $P^{x_i}_i$ is the law on  $C(\R_+,\mathbb{R})$ of the one-dimensional SDE (\ref{freedyn1_dim}) with initial condition $x_i \in \R$.\\
We denote by  $p_t(x_i,\cdot) $ its density function at time $t$ with respect to $m$:
\begin{equation}
 P^{x_i}_i \circ X(t)^{-1}(dy_i) = p_t(x_i,y_i)\, m(dy_i) .
\end{equation}

\subsection{A finite-dimensional approximation}\label{sec:laws}

As usual, we approximate the infinite-volume dynamics by a sequence of finite-volume dynamics.\\
Let $\Lambda$ be a finite subset of $\Z^d$, and define 
\begin{equation}
\Lambda^- = \lbrace i \in \Lambda: \lbrace i + \mathcal{N} \rbrace \subset \Lambda \rbrace
\end{equation}
its $\mathcal{N}$-interior.\\
Let  $Q^{x}_{\Lambda}$ denotes the law of the finite-volume dynamics
\begin{equation}
 \begin{cases}
& dX_i(t) =dB_i(t) + \big( -\frac{1}{2}U^{\prime}(X_i(t)) + \beta \, b_i(t,X) \big)\, dt, \,  i \in \Lambda^-\\
& dX_i(t) =dB_i(t) -\frac{1}{2}U^{\prime}(X_i(t)) \, dt,  \,  i \in \Lambda \setminus\Lambda^-\\
& X_{\Lambda}(0) = x_{\Lambda}.\label{finiteSDE}
\end{cases}
\end{equation}
It is a perturbation of the finite-volume free dynamics
$
P^x_\Lambda=\bigotimes_{i \in \Lambda} P^{x_i}_i .
$


\subsection{Cluster expansion of the finite-dimensional density}

First we expand the finite-volume density of the perturbed system w.r.t.  the free system.
\begin{LEM}
At any time $t$, $Q_{\Lambda}^{x}\circ X(t)^{-1}$ is absolutely continuous with respect to $P_{\Lambda}^{x}\circ X(t)^{-1}$ on
$\mathbb{R}^{\Lambda}$ and its density is given by
\begin{equation}\label{RN2}
f^t_{\Lambda}(x,y):=\frac{dQ_{\Lambda}^{x} \circ X(t)^{-1}}{dP_{\Lambda}^{x} \circ X(t)^{-1}} ( y_{\Lambda})  =
\E_{P_{\Lambda, [0,t]}^{xy}} \biggl [ \exp \biggl ( - \sum_{A \subset \Lambda} \Psi_{A, [0,t]}(X) \biggr )
 \biggr ] .
\end{equation}
 $P^{xy}_{[0,t]}$ denotes the law of the bridge on $[0,t]$ 
obtained by conditioning $P_\Lambda$ to be at time
$0$ in $x_\Lambda$ and at time $t$ in $y_\Lambda$, and
the  functional $\Psi_{A,[0,t]}$ satisfies
\begin{equation} \label{Psi}
 \Psi_{A,[0,t]}(X) = \begin{cases}
              -\beta \int_0^t b_i(s,X)d \overline{B}_i(s) + \frac{\beta^2}{2} \int_0^t b_i^2(s,X)ds & \text{ if } \exists i : A= \mathcal{N} + i \\
               0 & \text{ otherwise }
 \end{cases}
\end{equation}
where the process  $\overline{B}$ is defined as
\begin{equation*}
\overline{B}_i(t)(\omega) = \omega_i(t) + \frac{1}{2}\int_0^t U^{\prime}(\omega_i(s)) ds.
\end{equation*}
\end{LEM}
\proof

By  Girsanov Theorem,  
\begin{equation*}
\begin{split}
dQ^x_{\Lambda}(X) &= \exp \biggl ( \sum_{i \in \Lambda^-} \Bigl (\beta \int_0^t b_i(s,X) d\overline{B}_i(s) 
-\frac{\beta^2}{2} \int_0^t b_i^2(s,X)ds \Bigr ) \biggr ) dP^x_{\Lambda}(X) \\
& =: M_{\Lambda,t}(X)\, dP^x_{\Lambda}(X)
\end{split}
\end{equation*}
Let $f$ be a bounded local function on $ \mathbb{R}^{\mathbb{Z}^d}$. Then
\begin{equation*}
\begin{split}
\mathbb{E}_{Q_{\Lambda}^{x}} (f(X(t))) & = \mathbb{E}_{P_{\Lambda}^{x}} \Big(M_{\Lambda,t}(X)f(X(t))\Big) \\
& = \int \mathbb{E}_{P_{\Lambda, [0,t]}^{xy}} \Big(M_{\Lambda,t}(X)f(X(t))\Big)  p_t(x_{\Lambda},y_{\Lambda}) \, m(dy_{\Lambda})\\
& = \int  f(y_{\Lambda}) \mathbb{E}_{P_{\Lambda, [0,t]}^{xy}} (M_{\Lambda,t}(X))  p_t(x_{\Lambda},y_{\Lambda}) \, m(dy_{\Lambda})
\end{split}
\end{equation*}
which leads to the desired result.
\qed
\begin{REM}
\begin{itemize}
 \item
The functional $\Psi$  is not defined a priori on the whole path space $\Omega$, but only for $\omega \in \Omega' \subset \Omega$ for which the
stochastic integral $\int_0^t b_i(s,\omega)d\omega_i(s)$ makes sense. 
\item
If we would assume the initial inverse temperature to be very small (i.e. $\beta_0$ vanishing), we could use the usual cluster expansion
techniques with respect to both $\beta_0$ and $\beta$ in space  to obtain a perturbative result around the free stationary
case ($\beta_0=\beta=0$). As we would like to treat the more general case where $\beta_0$ is not necessarily close to 0, we now develop  a more involved
space-time cluster expansion technique, which allows us to control space and time simultaneously. 
\end{itemize}
\end{REM}


\medskip

In the following let us perform, for a fixed time $t$, the cluster expansion for $f^t_{\Lambda}(x,y)$ w.r.t. the intensity $\beta$ of the
dynamical perturbation.\\

We decompose the time interval $[0,t]$ into 
 $M$ subintervals $I_j:=[jT,(j+1)T]$ with length $T=t/M$,
where $T$ is a time step length
larger than the range $t_0$ of the time-memory of the drift $b$, in such a way that
$[jT-t_0,jT+t_0] \subset [(j-1)T,(j+1)T].$ This latter condition is important to control the range of the time interaction. \\
A \textit{temporal edge} is a unit space-time pair of the form
$(i,I_j)$ with $i \in \mathbb{Z}^d$ and $j \in \mathbb{N}$. Its vertices are the points $(i,jT)$ and $(i,(j+1)T)$ in $\mathbb{Z}^d \times \mathbb{R}_+$.
A \textit{space cluster} $\gamma^j$, $j \in \mathbb{N}$ is a finite collection of pairwise space-connected temporal edges, that is 
$\gamma^j = \{ (i_1,I_j),...,(i_m,I_j)\}, i_1,...,i_m \in \Z^d,$
where the sequence of subsets  $i_1 +\mathcal{N},   i_2 + \mathcal{N},\cdots, i_m + \mathcal{N} $ is connected:
$$
(i_1 + \mathcal{N})\cap( i_2  + \mathcal{N}) \neq
\emptyset, \dots, (i_{m-1} + \mathcal{N})\cap( i_m + \mathcal{N}) \neq \emptyset.
$$
Two space clusters $\gamma_1^j$ and $\gamma_2^j$ are called \textit{compatible}
if no temporal edge of the first one is space-connected with one temporal edge of the other one.\\
A \textit{time cluster} $\tau^i$, $i \in \Z^d$, is a finite collection of temporal edges of the following type
\begin{equation*}
\tau^i = \{ (i,I_j),...,(i,I_{j+r})\}, j,r \in \mathbb{N}.
\end{equation*}
We call \textit{space-time cluster} $\Gamma$ a non-empty collection of space and time clusters of the form
\begin{equation*}
\Gamma = \{ \gamma_1^{j_1},...,\gamma_s^{j_s};\tau_1^{i_1},...,\tau_p^{i_p}\}.
\end{equation*}
The \textit{spatial support} of $\Gamma$ is the set denoted by $[\Gamma]$ of all vertices belonging to the temporal edges which compose $\Gamma$. 
We denote by $[\Gamma]_{k,l}$ the set of all  vertices belonging to the temporal edges which compose $\Gamma$ except $j=k$ and $j=l$.
Two space-time clusters are called \textit{non-intersecting} if their space clusters are compatible and their time clusters are disjoint. 

\begin{PRP} \label{CEofft}
There exist cluster weights $K^t_{\Gamma}(x,y)$ indexed by space-time clusters $\Gamma \subset \Lambda \times [0,t]$, which depend on
$t,\beta,x$ and $y$ such that
\begin{equation} \label{eq:CEofft}
f^t_{\Lambda}(x,y) = 1 + \sum_{v \in \N^*} \sum_{\{\Gamma_1,...,\Gamma_v\}} K^t_{\Gamma_1}(x,y)\cdots K^t_{\Gamma_v}(x,y)
\end{equation}
where the last summation is on all pairwise non-intersecting space-time clusters $\Gamma_l$ included in $\Lambda \times [0,t]$.
\end{PRP}
\proof
 By simplicity, denote $\Psi_{k,j}$ instead of $\Psi_{k+\mathcal{N},I_j}$. We  develop \eqref{RN2} decomposing the bridge $ {P_{\Lambda, [0,t]}^{xy}}$ on
the time interval $[0,T]$ into a concatenation of bridges of the form $P_{\Lambda, I_j}^{x^{(j)}x^{(j+1)}}$:
\begin{eqnarray*}
&&f^t_{\Lambda}(x,y)   = 
\E_{P_{\Lambda, [0,t]}^{xy}} \biggl [ \exp \biggl ( - \sum_{k \in \Lambda^-} \Psi_{k + \mathcal{N}, [0,t]}(X) \biggr )
 \biggr ] \\
&&= \int\int \prod_{j=0}^{M-1} \prod_{k \in \Lambda^-} e^{ - \Psi_{k,j}(X)}
 \underset{0\leq j\leq M-1}\otimes P_{\Lambda, I_j}^{x^{(j)}x^{(j+1)}}(dX) 
  \underset{{ i \in \Lambda \atop 0 \leq j\leq M-2}} \prod p_{T}(x_i^{(j)}, x_i^{(j+1)}) \underset{{i \in \Lambda \atop 0 \leq j\leq M-2 }}\otimes
m(dx_i^{(j+1)}) \\
&& =\int \prod_{j=0}^{M-1} \int \prod_{k \in \Lambda^-}   e^{ - \Psi_{k,j}(X)}
  P_{\Lambda, I_{j-1}}^{x^{(j-1)}x^{(j)}}(dX) P_{\Lambda, I_j}^{x^{(j)}x^{(j+1)}}(dX) 
  \underset{{i \in \Lambda \atop 0 \leq j\leq M-2}} \prod
p_{T}(x_i^{(j)}, x_i^{(j+1)}) \underset{{i \in \Lambda \atop 0 \leq j\leq M-2 }} \otimes m(dx_i^{(j+1)})
x_i^{(j+1)}) \underset{{i \in \Lambda \atop 0 \leq j\leq M-2 }} \otimes m(dx_i^{(j+1)}) 
\end{eqnarray*}
where $x^{(0)}:=x \in \R^{\Lambda}$ and $x^{(M)}:=y \in \R^{\Lambda}$. Now use
\begin{equation*}
\begin{split}
\prod_{k \in \Lambda^-}   e^{ - \Psi_{k,j}(X)} & =  \prod_{k \in \Lambda^-}  (1+ e^{ - \Psi_{k,j}(X)} -1) \\
& = 1+ \sum_{n\geq 1}\sum_{\{\gamma_1^j,...,\gamma_n^j\}} \prod_{l=1}^n\prod_{(k,I_j)\in\gamma_l^j} (e^{-\Psi_{k,j}(X)-1})
\end{split}
\end{equation*}
where the last summation is over all pairwise compatible space clusters included in $\Lambda \times [0,t]$. \\
On the other hand for $z^{(0)},...,z^{(M)} \in \R$,
\begin{equation}
\begin{split}
\prod_{j=0}^{M-2} p_{T}(z^{(j)}, z^{(j+1)}) & =  \prod_{j=0}^{M-2} (1+ p_{T}(z^{(j)}, z^{(j+1)}) -1 ) \\
& = 1 + \sum_{\tau} \prod_{I_j \in \tau}  (p_{T}(z^{(j)}, z^{(j+1)}) -1) \\
& = 1 + \sum_{p\geq 1} \sum_{\{\tau_1,...,\tau_p\}} \prod_{u=1}^p \prod_{I_{j} \in \tau_{u}}
( p_{T}(z^{(j)}, z^{(j+1)}) -1 )
\end{split}
\end{equation}
where the summation on the second line is over all collections $\tau$ of time intervals of the type $I_j \subset [0,t]$ and 
the last summation on the third line is over all pairwise disjoint collections of such consecutive time intervals. One obtains
\begin{equation}
\begin{split}
f^t_{\Lambda}(x,y)  &= \\
\int_{\mathbb{R}^{|\Lambda|(M-1)}} & \prod_{j=0}^{M-1} \int_{\mathbb{R}^{|\Lambda|}} 
\biggl ( 1 + \sum_{n \geq 1} \sum_{\lbrace \gamma^j_1,...,\gamma^j_n \rbrace}  \prod_{l=1}^n \prod_{(k ,I_j)\in \gamma^j_l}
\bigl ( e^{-\Psi_{k, j}(X)} -1 \bigr ) \biggr )P_{\Lambda, I_{j-1}}^{x^{(j-1)}x^{(j)}}(dX) P_{\Lambda, I_j}^{x^{(j)}x^{(j+1)}}(dX) \\
& \prod_{i \in \Lambda} \biggl ( 1 + \sum_{p \geq 1} \sum_{\{\tau^i_1,...,\tau^i_p\}} \prod_{u=1}^p \prod_{I_{j} \in \tau^i_{u}}( p_{T}(x_i^{(j)}, x_i^{(j+1)})
-1 )
\biggr )
\underset{{i \in \Lambda \atop 0 \leq j\leq M-1 }} \otimes m(dx_i^{(j)}) \\
& =: 1 + \sum_{v \geq 1} \sum_{\{\Gamma_1,...,\Gamma_v\}} K^t_{\Gamma_1}(x,y)\cdot...\cdot K^t_{\Gamma_v}(x,y) 
\end{split}
\end{equation}
where the last summation is on all pairwise non-intersecting space-time clusters  $\Gamma_l$ included in $\Lambda \times [0,t]$. 
Therefore, for $\Gamma= \{ \gamma_1^{j_1},...,\gamma_s^{j_s};\tau_1^{i_1},...,\tau_p^{i_p}\}$, the cluster weight $K^t_{\Gamma}$ is defined by
\begin{eqnarray}\label{clusterest}
 K^t_{\Gamma}(x,y) & =& \int  \prod_{m=1}^s \int \prod_{k \in \gamma_m^{j_m}}\bigl ( e^{-\Psi_{k, j_m}(X)} -1 \bigr ) P_{\Lambda,
I_{j_m-1}}^{x^{(j_m-1)}x^{(j_m)}}(dX) P_{\Lambda, I_{j_m}}^{x^{(j_m)}x^{(j_m+1)}}(dX) \nonumber \\
&& \prod_{u=1}^p \prod_{I_j \in \tau_u^{i_u}}( p_{T}(x_{i_u}^{(j)}, x_{i_u}^{(j+1)}) -1 )\otimes_{(i,j) \in [\Gamma]_{0,M}}m(dx_i^{(j)}) \nonumber\\
& =& \int \prod_{m=1}^s \mathcal{K}(\gamma_m^{j_m}) \prod_{u=1}^p \prod_{I_j \in \tau_u^{i_u}} ( p_{T}(x_{i_u}^{(j)}, x_{i_u}^{(j+1)}) -1 )\otimes_{(i,j) \in
[\Gamma]_{0,M}}m(dx_i^{(j)})
\end{eqnarray}
with, for any $2\leq j\leq M-2$,
\begin{equation*}
\mathcal{K}(\gamma^j):= \int \prod_{k \in \gamma^j} (e^{-\Psi_{k,j}(X)}-1)\otimes_{i \in \Lambda}
P_{i,I_{j-1}}^{x_i^{(j-1)}x_i^{(j)}}(dX_i)P_{i,I_{j}}^{x_i^{(j)}x_i^{(j+1)}}(dX_i) \label{Kj}
 \end{equation*}
and for the space-cluster $\gamma^1$, taking into account the fixed boundary condition $x^{(0)}=x$
\begin{equation*}
\mathcal{K}(\gamma^1) := \int \prod_{k \in \gamma^1} (e^{-\Psi_{k,1}(X)}-1)\otimes_{i \in \Lambda} 
P_{i,I_{0}}^{x_ix_i^{(1)}}(dX_i)P_{i,I_{1}}^{x_i^{(1)}x_i^{(2)}}(dX_i), \label{K0}
\end{equation*}
resp. for the space-cluster $\gamma^{M-1}$ on the time interval $I_{M-1}=[t-T,t]$, taking into account the fixed boundary condition $x^{(M)}=y$
\begin{equation*}
\mathcal{K}(\gamma^{M-1}) := \int \prod_{k \in \gamma^{M-1}} (e^{-\Psi_{k,M-1}(X)}-1)\otimes_{i \in \Lambda}
P_{i,I_{M-2}}^{x_i^{(M-2)}x_i^{(M-1)}}(dX_i)P_{i,I_{M-1}}^{x_i^{(M-1)}y_i}(dX_i).\label{KM}
\end{equation*}
\qed

\subsection{Cluster Estimates}
The next step is to estimate the cluster weights  $K^t_{\Gamma}(x,y)$, defined by \eqref{clusterest}, as function of the small parameter $\beta$.
\begin{PRP}\label{cluster estimates}
Let $\Gamma=\{ \gamma_1^{j_1},...,\gamma_{s}^{j_s};\tau_1^{i_1},...,\tau_p^{i_p}\}$ be a space-time cluster.
There exists a function $\lambda(\beta)>0$ 
vanishing when $\beta$ tends to 0
such that the cluster weight $K^t_{\Gamma}(x,y)$  is bounded uniformly in time and space as follows:
\begin{equation} \label{lambdabeta}
\sup_{t \geq 0}\sup_{x,y} |K^t_{\Gamma}(x,y)| \leq \lambda(\beta)^{|\Gamma|}
\end{equation}
where $|\Gamma|$, the cardinality of  $\Gamma$, is the total number of unit temporal edges which compose $\Gamma$.
\end{PRP}
\proof
To bound  the cluster weights we need to interchange integration and products in \eqref{clusterest}. Therefore we make use of the following inequalities, 
generalizing
H\"older inequalities, see \cite{MVZ00} Lemma 5.2.
\begin{LEM}\label{Minlos}
Let $(\mu_z)_{z\in \chi}$ be a family of probability measures, each one defined on a measurable space $E_z$ where the elements $z$ belong to some finite set
$\chi$. Let $(g_k)_k$ be a family of functions on $E_{\chi} = \times_{z \in \chi}E_z$ such that each $g_k$ is $\chi_k$-local for a certain $\chi_k \subset \chi$
in the sense that 
\begin{equation}
 \forall e  \in E_{\chi}, \quad g_k(e) = g_k(e_{|\chi_k}) ,
\end{equation}
and let $(\rho_k)_k$ be positive numbers such that, for all $z \in \chi$, ${\displaystyle \sum_{\{k: \chi_k \ni z\}} 1/\rho_k \leq 1}$.
Then
\begin{equation}
\biggl | \int_{E_{\chi}} \prod_{k} g_k \otimes_{z \in \chi} d\mu_z \biggr | \leq \prod \biggl( \int_{E_{\chi_k} } |g_k|^{\rho_k} \otimes_{z \in \chi_k}
d\mu_z\biggr)^{1/\rho_k}.
\end{equation}
\end{LEM}
 
 \medskip

We apply lemma \ref{Minlos} with $\chi := \gamma^j + \mathcal{N}$, $\chi_k := k + \mathcal{N}$, $E_z:= C(\mathbb{R}_+,\mathbb{R})$, $g_k:=e^{-\Psi_{k,j}}-1$, 
$\mu_k := P_{k,I_{j-1}}^{x_k^{(j-1)}x_k^{(j)}} \otimes P_{k,I_j}^{x_i^{(j)}x_k^{(j+1)}}$ and  $\rho_i = 4|\mathcal{N}|$ for all $i$.
Since for each $i \in \Lambda$, there is at most $|\mathcal{N}|$ factors $k$ such that $\Psi_{k,j}(X)$ depends on $X_i$, the assumption $\sum_{k + \mathcal{N}
\ni i} \frac{1}{ 4|\mathcal{N}|} \leq 1$ is satisfied.\\
We then obtain the upper bound
\begin{equation}
\begin{split}
|\mathcal{K}(\gamma^j)| & \leq \prod_{k \in \gamma^j} \biggl [\int (e^{-\Psi_{k,j}}-1)^{4N} 
\otimes_{i \in k+\mathcal{N}}P_{i,I_{j-1}}^{x_i^{(j-1)},x_i^{(j)}}(dX_i)P_{i,I_j}^{x_i^{(j)},x_i^{(j+1)}}(dX_i) \biggr]^{1/4|\mathcal{N}|}\\
& =: \prod_{k \in \gamma^j}  \mathbb{K}_{k,j}(x^{(j-1)},x^{(j)},x^{(j+1)}) .
\end{split}
\end{equation}

Remark at that place that  we especially used the space-locality of the interaction $b$ (assumption (B2)).\\
Therefore
\begin{equation}
|K^t_{\Gamma}(x,y)| \leq \int \prod_{m=1}^s\prod_{k \in \gamma_{m}^{j_m}}  \mathbb{K}_{k,j}(x^{(j-1)},x^{(j)},x^{(j+1)}) 
\prod_{u=1}^p \prod_{I_j \in \tau_u^{i_u}} ( p_{T}(x_{i_u}^{(j)}, x_{i_u}^{(j+1)}) -1 )\otimes_{(i,j) \in [\Gamma]_{0,M}}m(dx_i^{(j)}). \label{KK}
\end{equation}
We apply once more lemma \ref{Minlos} to bound the right hand side of \eqref{KK} by
\begin{equation*}
 \prod_{m=1}^s\prod_{k \in \gamma_m^{j_m}} \biggl ( \int \mathbb{K}^{N_1}_{k,j}(x^{(j-1)},x^{(j)},x^{(j+1)}) \otimes_{(i,j)} m(dx_i^{(j)}) \biggr )^{1/N_1}
\prod_{u=1}^p\prod_{I_j \in \tau_u^{i_u}} \biggl ( \int (p_{T}(x_{i_u}^{(j)}, x_{i_u}^{(j+1)}) -1)^{N_2} \otimes_{(i,j)} m(dx_i^{(j)})\biggr )^{1/N_2}
\end{equation*}
for any right choice of $N_1,N_2$ satisfying $2|\mathcal{N}|/N_1+ 2/N_2 \leq 1$. Choose e.g. $N_1=4|\mathcal{N}|$ and $N_2=4$. 
In the two next lemmas we will show
that the first integral describing the spatial interaction (resp. the second integral describing the time interaction) is bounded uniformly in $t, x$ and $y$ by
a function $C_1(\beta)$ (resp. by $C_2(\beta)$), which leads to
\begin{equation} \label{eq:KtGamma}
|K^t_{\Gamma}(x,y)|\leq C_1(\beta)^{\sum_m |\gamma_m^{j_m}|} C_2(\beta)^{ \sum_u |\tau_u^{i_u}|} \leq \max(C_1,C_2)(\beta)^{\sum_m |\gamma_m^{j_m}| + 
\sum_u |\tau_u^{i_u}|}
\end{equation}
which yields the claim (\ref{lambdabeta}) with  $\lambda(\beta):= \max(C_1,C_2)(\beta) $.\\
In the next lemma we prove appropriate upper bounds for the spatial interaction, that is for the integral of $\mathbb{K}$, treating first the case where the 
space cluster $\gamma^j$  does not contain any boundary temporal edge, that is $j \not = 0$ and $j \not =M$ .
\begin{LEM} \label{lem:C1}
Let $j=1,..,M-1$. There exists a positive real number $C_1$ depending only on $\beta$ (and uniform in $t,x,y,k$ and j), vanishing when $\beta$ goes to 0, such
that the following upper bound holds
\begin{equation} \label{C1}
\int \mathbb{K}^{4|\mathcal{N}|}_{k,j}(x^{(j-1)},x^{(j)},x^{(j+1)}) \otimes_{i \in k + \mathcal{N}} m(dx_i^{(j)}) \leq C_1(\beta)^{4|\mathcal{N}|}.
\end{equation}
\end{LEM}
\proof
Let fix $k$. Then 
\begin{equation*}
\begin{split}
&  \int \mathbb{K}^{4|\mathcal{N}|}_{k,j}(x^{(j-1)},x^{(j)},x^{(j+1)}) \otimes_{i \in k + \mathcal{N}} m(dx_i^{(j)}) \\
 & \leq \int \int (e^{-\Psi_{k,j}(X)}-1)^{4|\mathcal{N}|} \otimes_{i \in k +\mathcal{N}}
P_{i,I_{j-1}}^{x_i^{(j-1)}x_i^{(j)}}(dX_i)P_{i,I_{j}}^{x_i^{(j)}x_i^{(j+1)}}(dX_i)m(dx_i^{(j-1)})m(dx_i^{(j)})m(dx_i^{(j+1)}) \\
& = \mathbb{E}_{P_{\Lambda}}\bigl( (e^{-\Psi_{k,j}(X)}-1)^{4|\mathcal{N}|}) \bigr).
\end{split} \label{M1}
\end{equation*}
We remark that, for any $\zeta \in \R$,
\begin{equation*}
(e^{\zeta}-1)^{4|\mathcal{N}|}  = \zeta^{4|\mathcal{N}|}\biggl (\int_0^1 e^{u\zeta}du\biggr)^{4|\mathcal{N}|} 
= \zeta^{4|\mathcal{N}|}\int_0^1...\int_0^1
e^{(u_1+...+u_{4|\mathcal{N}|})\zeta} du_1...du_{4|\mathcal{N}|}. \label{FIntegral}
\end{equation*}
Hence
\begin{equation*}
\mathbb{E}_{P_{\Lambda}}\bigl( (e^{-\Psi_{k,j}(X)}-1)^{4|\mathcal{N}|}) \bigr) =
\int_{[0,1]^{4|\mathcal{N}|}}
\mathbb{E}_{P_{\Lambda}}(\Psi_{k,j}^{4|\mathcal{N}|} e^{-(u_1+...+u_{4|\mathcal{N}|})\Psi_{k,j}})\, du_1...du_{4|\mathcal{N}|}.
\end{equation*}
The expectation above can be written as $\frac{\partial}{\partial z^{4|\mathcal{N}|}}
\mathbb{E}_{P_{\Lambda}}(e^{-z\Psi_{k,j}})\biggr|_{z=u_1+...+u_{4|\mathcal{N}|}}$, the $4|\mathcal{N}|^{th}$-derivative of the Laplace transform $L$ of the
functional $\Psi_{k,j}$ at $z=u_1+...+u_{4|\mathcal{N}|}$.
Let us analyse $L$:
\begin{equation*}
\begin{split}
L(z) & = \mathbb{E}_{P_{\Lambda}}(e^{-z\Psi_{k,j}})\\
& = \mathbb{E}_{P_{\Lambda}}\biggl ( \exp \biggl [ z \beta\int_{I_j}b_k(s,X) d \overline{B}_k(s) - z^2 \beta^2\int_{I_j} b_k^2(s,X) ds\biggr ] \exp \biggl [
 z \biggl( z -  \frac{1}{2} \biggr)\beta^2 \int_{I_j} b^2_k(s,X) ds \biggr]\biggr) \\
& \leq \mathbb{E}_{P_{\Lambda}}^{1/2} \biggl ( \exp \biggl[ 2 z\beta  \int_{I_j} b_k(s,X) d \overline{B}_k(s) - \frac{(2 z\beta)^2}{2}\int_{I_j}b_k^2(s,X)ds
\biggr]\biggr )\mathbb{E}_{P_{\Lambda}}^{1/2} \biggl ( \exp \biggl[ z(2z-1)\beta^2 \int_{I_j}b^2_k(s,X)ds\biggr]\biggr )  \\
& =  \mathbb{E}_{P_{\Lambda}}^{1/2} \biggl ( \exp \biggl[ z(2z-1)\beta^2\int_{I_j}b^2_k(s,X) ds\biggr]\biggr )
\end{split}
\end{equation*} 
due to the $P_{\Lambda}$-martingale property of
$t \mapsto \exp \bigl[ 2 z\beta  \int_{jT}^t b_k(s,X) d \overline{B}_k(s) - \frac{(2 z\beta)^2}{2}\int_{jT}^t b_k^2(s,X)ds \bigr]$.
To bound not only $L$ but its derivatives, we extend it to the complex plane and notice that 
\begin{equation}
\biggl | \frac{\partial}{\partial z^{4|\mathcal{N}|}} L(z) \biggr| \leq \frac{4|\mathcal{N}|!}{\rho^{4|\mathcal{N}|}} \sup_{\{\zeta \in \C:|\zeta-z|=\rho\}}
|L(\zeta)|\label{supL}
\end{equation}
as soon as $L$ is well defined on $B(z,\rho)=\{\zeta \in \C:|\zeta-z|\leq\rho \}$. On $B(z,\rho)$ one has
\begin{equation*}
\begin{split}
\biggl| \exp \biggl[  \zeta (2\zeta - 1) \beta^2\int_{I_j} b_k^2(s,X) ds \biggr] \biggr| & \leq \exp \biggl[ {\mathcal Re}(2\zeta^2-\zeta) \beta^2\int_{I_j}
b_k^2(s,X) ds \biggr] \\
& \leq \exp \biggl[3(\rho \beta)^2  \int_{I_j} b_k^2(s,X) ds \biggr]\\
& \leq \exp \biggl ( 3 (\rho \beta)^2 T \bar b^2 \biggr ). 
\end{split}
\end{equation*}
Therefore (\ref{supL}) becomes 
\begin{equation*}
\biggl | \frac{\partial}{\partial z^{4|\mathcal{N}|}} L(z) \biggr| \leq \frac{4|\mathcal{N}|!}{\rho^{4|\mathcal{N}|}}  \exp (3 (\rho \beta)^2 T \bar b^2).
\end{equation*}
We minimize the r.h.s.  choosing $\rho^2=\frac{2|\mathcal{N}|}{3 T\beta^2 \bar b^2}$.  Thus
$
\biggl | \frac{\partial}{\partial z^{4|\mathcal{N}|}} L(z) \biggr| \leq c \, (\beta^2 T )^{2|\mathcal{N}|}
$
where $c$ is a positive constant depending only on $\bar b$ and $|\mathcal{N}|$. Taking the time step $T$ of the order of $1/\beta$, this leads to the desired 
inequality (\ref{C1}) with $C_1(\beta):= \sqrt{c}\sqrt{\beta}$. 
\qed

Let us short comment how to compute a similar upper  bound in the case of $j=0$ (resp. in the case $j=M$, in a symmetric way).
In that case the spatial support of the cluster $\gamma^0$ (resp. $\gamma^M$) contains the vertex $x$ (resp. y).
In that case one space boundary is fixed (equal to $x$ or $y$) and we have to control integrals of the type
$$
\int \int (e^{-\Psi_{k,0}(X)}-1)^{4|\mathcal{N}|} \otimes_{i \in k +\mathcal{N}}
P_{i,I_{0}}^{x_ix_i^{(1)}}(dX_i)m(dx_i^{(1)}) = 
\mathbb{E}_{P^x_{\Lambda}}\bigl( (e^{-\Psi_{k,0}(X)}-1)^{4|\mathcal{N}|}) \bigr).
$$
We then can use the same arguments as in Lemma \ref{lem:C1}, 
that is identify an exponential martingale and make use of the boundedness of the drift $b$. 

To estimate the  time interaction upper bound  $C_2$ appearing 
in \eqref{eq:KtGamma}, i.e. the forth moment of the transition
kernel $p_t$ of the one-dimensional free dynamics, we also have to distinguish between different types of time clusters composing the space-time cluster
$\Gamma$: those containing a boundary temporal edge $I_0$ or $I_M$ and the other time clusters. The next lemma provides an upper bound in that latter case.
\begin{LEM}
There exists positive constants $c',c'' $ depending only on the self potential $U$ such that
\begin{equation} \label{eq:C2}
\Big( \int (p_{1/\beta}(z, z') -1)^{4}  m(dz)m(dz') \Big)^\frac{1}{4} \leq c' e^{-\frac{c'' }{\beta}}.
\end{equation}
\end{LEM}
\proof
First 
\begin{equation*}
\begin{split}
\int (p_T(z, z') -1)^{4}m(dz)m(dz') & \leq \int ||p_T(\cdot,\cdot)-1||^4_{L^{\infty}}\, m(dz)m(dz') = 
||p_T(\cdot,\cdot)-1||^4_{L^{\infty}}.
\end{split}
\end{equation*}
Now, under the ultracontractivity assumption (B1) on the self-interaction  $U$, one has a uniform  exponential convergence  of $p_T$ to 1 (see e.g. 
the details of the proof in the appendix of \cite{DPRZ02}). Moreover the rate of convergence is equal to the spectral gap of $p_T$. Thus 
\begin{equation}
\exists c', c''>0, \forall T>0, \quad ||p_T(\cdot,\cdot)-1||^4_{L^{\infty}} \leq c' e^{-c''T}
\end{equation}
where $c''$ is the spectral gap of $(p_t)_t$.
We obtain the claim \eqref{eq:C2} taking $T=1/\beta$ .
\qed

When the time cluster $\tau$ composing $\Gamma$ contains the boundary temporal edge  $I_0$ (resp. $I_{M-1}$) one has to estimate the simple integral 
$\int (p_{T}(x, z) -1)^{4}  m(dz) $ (resp. $\int (p_{T}(z, y) -1)^{4}  m(dz) $ ) instead of the above double integration \eqref{eq:C2} under $m\otimes m$. It
 vanishes with an exponential rate  uniformly in $x$ and $y$ when $T$ tends to
infinity. \\
Therefore one can take in \eqref{eq:KtGamma} the upper bound $C_2(\beta):= c' e^{-\frac{c'' }{\beta}}$.

\subsection{Cluster expansion and estimates of the  logarithm of the finite-dimensional density}

To complete Proposition \ref{CEofft} we are now computing an expansion of the logarithm of the density at time $t$ of the finite-dimensional SDE
\eqref{finiteSDE}. 
\begin{PRP}
For $\beta $ small enough, the logarithm of the Radon-Nikodym derivative \eqref{RN2} expands as
\begin{equation}
\log f^t_{\Lambda}(x,y)= -\sum_{\Delta \subset \Lambda} \Phi^{t}_{\Delta}(x,y)
\end{equation}
with
\begin{equation} \label{eq:PhibetaDelta}
\Phi^t_{\Delta}(x,y) =  \sum_{n\geq 0}\sum_{{\{\Gamma_1,..,\Gamma_n\}\atop
Tr(\Gamma_1,...,\Gamma_n)=\Delta}}C(\Gamma_1,...,\Gamma_n) \, \mathcal{K}^{x,y}(\Gamma_1)
\cdots \mathcal{K}^{x,y}(\Gamma_n)
\end{equation}
where the second sum runs over all collections of disjoint space-time clusters such that their union is connected and $C(\Gamma_1,...,\Gamma_n)$ are
purely combinatorial coefficients independent of $x$ and $y$.  
\end{PRP}
\proof
We alread know that the density function \eqref{RN2} decomposes as 
\begin{equation*}
f^t_{\Lambda}(x,y)=
\E_{P_{\Lambda, [0,t]}^{x,y}} \biggl [ \exp \biggl ( - \sum_{A \subset \Lambda} \Psi_{A, [0,t]}(X) \biggr )
 \biggr ] , \nonumber
\end{equation*}
which expands as in \eqref{eq:CEofft} with cluster weights of the form $K^t_{\Gamma}(x,y)$.
We now use Koteck\'y and Preiss criterion proven in \cite{KP86} to derive an expansion of its logarithm.\\
Let $\Gamma$ be a space-time cluster. We say that another space-time cluster $\Gamma'$ is \textit{incompatible} with $\Gamma$ if their
associated supports intersect, and we denote this property   by the symbol $\Gamma \nsim \Gamma'$. 
Take now  $\bar \beta$ small enough such that for $\beta \leq
\bar \beta$,
\begin{equation}
\begin{split} \label{cluwei}
\sup_{x,y \in \mathbb{R}} \sup_{t>0} \sum_{\Gamma' \nsim \Gamma} |K^t_{\Gamma'}(x,y)| e^{|\Gamma'| + \log(|\Gamma'|)} 
\Big(\leq \sum_{\Gamma' \nsim
\Gamma}|\Gamma'| (\lambda(\beta)e)^{|\Gamma'|} \Big) \leq |\Gamma|  .
\end{split}
\end{equation}
So, following assertion (2) in \cite{KP86}
the logarithm of $f^t_{\Lambda}(x,y)$ is expandable, and the following holds:
\begin{equation} \label{lnft}
\ln(f^t_{\Lambda}(x,y)) = \sum_{n\geq 0}\sum_{\Gamma_1,..,\Gamma_n } C(\Gamma_1,...,\Gamma_n)\,
\mathcal{K}^{x,y}(\Gamma_1)\cdot...\cdot \mathcal{K}^{x,y}(\Gamma_n).
\end{equation}
The second sum runs over collections of compatible space-time clusters such that their union is connected and  $C(\Gamma_1,...,\Gamma_n)$ are combinatorial
coefficients coming from the Taylor expansion. Let us now order the space-time clusters in terms of  their spatial  projections, which are subsets of $\Lambda$:
 If $Tr$ denotes the projection on the spatial support we  rewrite \eqref{lnft} as $-\sum_{\Delta \subset \Lambda} \Phi^t_{\Delta}(x,y) $ 
where $\Phi^t_{\Delta}$ is an interaction function given by \eqref{eq:PhibetaDelta}. \\
Moreover $\Phi^t_{\Delta}$ 
is $\mathcal{F}_{\Delta} \times \mathcal{F}_{\Delta}$-measurable since the cluster weights $\mathcal{K}^{x,y}(\Gamma)$ depend on $x$ on
$supp(\Gamma)\cap(\mathbb{Z}^d \times \{0 \})$ and on $y$ on $supp(\Gamma)\cap(\mathbb{Z}^d
\times \{t \})$ whose traces are included in $\Delta$.

\qed

Moreover, Koteck\'y and Preiss provide a useful estimate of the convergence rate of the interaction function $\Phi^t_{\Delta}$ in terms of $\Delta$, see
inequality (4) in \cite{KP86}:  
\begin{LEM}\label{lemmaSDC}
The function $\Phi^t_{\Delta}$ satisfies
\begin{equation}
\lim_{\beta \rightarrow 0} \sup_{i \in \mathbb{Z}^d} \sup_{t > 0} \sum_{\Delta \ni i} (|\Delta|-1)||\Phi^t_{\Delta}||_{\infty} = 0 . \label{eq:SDC}
\end{equation}
\end{LEM}
\proof
 Indeed Koteck\'y and Preiss proved the following bound for the interaction function:
\begin{equation*}
\sup_{i \in \mathbb{Z}^d} \sup_{t > 0} \sum_{\Delta \ni i} (|\Delta|-1)||\Phi^t_{\Delta}||_{\infty} \leq 1.
\end{equation*}
Therefore, since the sum on $\Delta$ converges uniformly in $i$ and $t$, we can interchange limit in $\beta$ and summation on $\Delta$ to obtain the
desired result \eqref{eq:SDC}. 

\qed

\subsection{Gibbsianness of the double-layer measure and Kozlov's representation theorem}\label{sec:doublelayer}

The rest of the proof of Theorem \ref{mainTh} follows the same structure as Steps 2 and 3 of  \cite{DR05}, Section 4, in which the drift $b$ is 
Markov and gradient.
Nevertheless, to make our paper
self-contained, we sketch the main arguments without giving as much details. \\
The time-evolved measure we are interested in, $Q^{\nu}\circ X(t)^{-1}$, is indeed a $\nu$-mixture of the measures $Q^{x}\circ X(t)^{-1}$ whose approximating
densities are $f^t_{\Lambda}(x,\cdot)$. Therefore, in order to prove the Gibbsianness of $Q^{\nu}\circ X(t)^{-1}$, we will prove as an intermediate step, the 
Gibbsianness of the so-called double-layer measure (or measure on the bi-space)
$\textbf{Q}^{\nu} := Q^{\nu}\circ (X(0),X(t))^{-1}$ defined  on the space $ \mathbb{R}^{\mathbb{Z}^d \times \{ 0,t\}}.$
\begin{LEM}
Let  $\nu\in \mathcal{G}_{\beta_0 }(\phi)$, where the interaction $\phi$ satisfies (A1). 
There exist an upper bound  $\bar \beta_0>0$ for the initial inverse temperature and an upper bound   $\bar \beta > 0$ for the intensity of the 
dynamical interaction such that, for $\beta_0 \leq \bar \beta_0$ and $\beta \leq \bar \beta$, the measure 
$\textbf{Q}^{\nu}$ is a 
Gibbs measure on the bi-space  
$\mathbb{R}^{\mathbb{Z}^d \times \{ 0,t\}}$ w.r.t. 
the a priori measure $m\otimes m$ with an interaction associated to  the 
Hamiltonian 
\begin{equation}\label{eq:Hamilbispace}
\textbf{H}_{(\Delta,\Delta')}(x,y) := h_{\Delta} - \sum_{i\in \Delta \cup \Delta'} \log(p_t(x_i,y_i)) + \sum_{A \subset \mathbb{Z}^d; A \cap (\Delta \cup
\Delta')\neq \emptyset} \Phi^t_{A}(x,y)
\end{equation}
where $h$ is the Hamiltonian function derived from $\phi$ and $(\Delta,\Delta')$ is short for $(\Delta \times \{ 0 \})\cup(\Delta' \times \{ t \})$.
\end{LEM}
\proof

Since the interaction $\phi$ of the initial Gibbs measure satisfies (A1), there exists $\bar \beta_0> 0$ such that
for $ \beta_0 \leq \bar \beta_0$,
\begin{equation} \label{eq:SDCt0}
\beta_0 \, \sup_{i \in \mathbb{Z}^d} \sum_{\Lambda \ni i} (|\Lambda| -1) ||\phi_{\Lambda}||_{\infty} < 1.
\end{equation}
This assumption implies Dobrushin's uniqueness condition, as it is proved e.g. in \cite{G88}, Proposition (8.8). 
In particular, for $ \beta_0 \leq \bar \beta_0$, $\mathcal{G}_{\beta_0 }(\phi)$ contains as unique element $\nu$, which  can be approximated e.g. by the
sequence of
finite-volume Gibbs measure $ \nu_{\Lambda}$ with free boundary condition. 


Since the sequence $Q^{\nu_{\Lambda}}_{\Lambda}$ converges towards $Q^{\nu}$ when $\Lambda$ increases to $\Z^d$, their joint projection at times $0$ and $t$
 converges  towards $\textbf{Q}^{\nu} := Q^{\nu} \circ (X(0),X(t))^{-1}$ on $ \mathbb{R}^{\mathbb{Z}^d \times \{ 0,t\}}$. 
$\textbf{Q}^{\nu}$ is Gibbs w.r.t. the a priori  measure $\textbf{m}(dx,dy)=p_t(x,y)m(dx)m(dy)$ and with interaction
\begin{equation}
\Psi_{\Delta}(x,y) := \phi_{\Delta}(x) + \Phi^t_{\Delta}(x,y), \quad  x,y \in \mathbb{R}^{\mathbb{Z}^d}, \Delta \subset \mathbb{Z}^d .\label{eq:Psi}
\end{equation}
It follows now from \eqref{eq:SDCt0} and \eqref{eq:Psi} that there exists a bound $\bar \beta$
for the intensity of the dynamical interaction such that, for any 
$\beta\leq \bar \beta$, the Dobrushin's uniqueness
assumption is satisfied for $\Psi$ on the bi-space. Therefore  $\textbf{Q}^{\nu}$ is 
 the  unique Gibbs measure on the bi-space associated to the interaction \eqref{eq:Psi} or, equivalently, 
 the  unique Gibbs measure associated to the Hamiltonian \eqref{eq:Hamilbispace} and the a priori measure $m\otimes m$.  
\qed

 Now the measure $\textbf{Q}^{\nu}$ can be easily desintegrated in a Gibbsian way  w.r.t. the finite dimensional projections at time $t$, 
 $\textbf{Q}^{\nu}(\cdot|X_{\Lambda^c}(t)=y_{\Lambda^c})$, which are defined for a.e. $y$. 

\begin{LEM}\label{lemma10}
Fix a finite set $\Lambda \subset \Z^d$. The conditional law of $Q^{\nu}\circ (X(0),X(t))^{-1}$ given $\{ X_{\Lambda^c}(t) = y_{\Lambda^c}\}$, denoted by
$\textbf{Q}^{\nu,y_{\Lambda^c}}$, 
is a Gibbs measure on $\mathbb{R}^{(\mathbb{Z}^d\times \{ 0 \})\cup (\Lambda \times \{ t \})}$ with reference measure $m$ and Hamiltonian
$\textbf{H}^{y_{\Lambda^c}}$ defined
by
\begin{equation}
\textbf{H}^{y_{\Lambda^c}}_{(\Delta,\Delta')}(x,z_{\Lambda}) = \textbf{H}_{(\Delta,\Delta')}(x,z_{\Lambda}y_{\Lambda^c}), \quad (\Delta,\Delta') \subset
\mathbb{Z}^d\times \Lambda.
\end{equation}
Furthermore $\textbf{Q}^{\nu,y_{\Lambda^c}}$ can be decoupled as follows
\begin{equation*}
\textbf{Q}^{\nu,y_{\Lambda^c}} (dx,dz_{\Lambda})  =
 \frac{1}{Z_{\Lambda}^{y_{\Lambda^c}}} \prod_{i \in \Lambda}p_t(x_i,z_i) \exp \biggl ( - \sum_{A\cap \Lambda \neq \emptyset}
\Phi^t_A(x,z_{\Lambda}y_{\Lambda^c})\biggr )m^{\otimes \Lambda}(dz_{\Lambda}) \tilde{Q}^{\nu,y_{\Lambda^c}}(dx)
\end{equation*}
where $\tilde{Q}^{\nu,y_{\Lambda^c}}(dx)$ is the unique Gibbs measure on $\mathbb{R}^{\mathbb{Z}^d}$ defined by the  interaction $\tilde{\Phi}^{y_{\Lambda^c}}$
given by
\begin{equation}
\begin{cases}
\tilde{\Phi}_i^{y_{\Lambda^c}} & =  \phi_i(x) - \1_{\{ i\in \Lambda^c\}}\log(p_t(x_i,y_i)), i \in \mathbb{Z}^d \\
\tilde{\Phi}_{\Delta}^{y_{\Lambda^c}} & =  \phi_{\Delta}(x) - \1_{\Delta \cap \Lambda = \emptyset}\Phi^t_{\Delta}(x,y_{\Lambda}^c), \Delta \subset \mathbb{Z}^d,
|\Delta| \geq 2 .
\end{cases}
\end{equation}
\end{LEM}
Indeed, due to the estimates already obtained, it is straightforward  to show that, for $\beta $ small enough, the interaction $\tilde{\Phi}^{y_{\Lambda^c}}$
satisfies
Dobrushin's uniqueness condition uniformly in $y$ and in $\Lambda$,  as perturbation of the initial interaction, see  \cite{DR05} Lemma 10 and Lemma 11.

\begin{LEM}\label{lem:step3}
The conditional law of $Q^{\nu}\circ X(t)^{-1}$ given $\{ X_{\Lambda^c}(t) = y_{\Lambda^c}\}$ admits a density w.r.t. $m^{\otimes \Lambda}(dz_\Lambda)$ given by
\begin{equation}
g^{t,y_{\Lambda^c}}_{\Lambda}(z_\Lambda) = \frac{1}{Z_{\Lambda}^{y_{\Lambda^c}}} 
\int_{\mathbb{R}^{\mathbb{Z}^d}} \prod_{i \in \Lambda}p_t(x_i,z_i) \exp \biggl
( - \sum_{A\cap \Lambda
\neq \emptyset} \Phi^t_A(x,z_\Lambda y_{\Lambda^c})\biggr ) \tilde{Q}^{\nu,y_{\Lambda^c}}(dx).
\end{equation}
Moreover this density is bounded from below and from above uniformly in $y$ and $t$, and it is quasilocal, i.e.
$$\lim_{\Delta \rightarrow \mathbb{Z}^d} \sup_{z,z': z_{\Delta}=z'_{\Delta}} |g^{t,z_{\Lambda^c}}_{\Lambda}(z_\Lambda) -
g^{t,z'_{\Lambda^c}}_{\Lambda}(z'_\Lambda)|=0.
$$
\end{LEM}
Boundedness and quasilocality of  $g^{t,y_{\Lambda^c}}_{\Lambda}$ allow to apply Kozlov's representation (Theorem 2 in \cite{K74}) which insures
the existence of an (absolute summable) interaction $\phi^t$ for $Q^{\nu}\circ X(t)^{-1}$.

\subsection{Additional remarks}\label{secAppl}
\subsubsection{Direct applications}
In this section we give some concrete examples for which the assumptions (B1)-(B3) on $U$ and $b$ are satisfied, and thus Theorem \ref{mainTh} and Corollary
\ref{mainCor} hold true. \\
Recall first some sufficient conditions which imply the ultracontractivity of the one-dimensional free dynamics (\ref{freedyn1_dim}), assumption (B1):
$$
(1) \liminf_{|x|\rightarrow \infty} U^{''}(x) > 0, \quad
(2) \quad \exists C \textrm{ s.t. } U^{''}-\frac{1}{2}(U^{'})^2 \leq C,\quad
(3) \quad \exists M > 0\textrm{  s.t.}\int_{|x|> M} \frac{1}{U^{'}(x)} dx < + \infty.
$$
Properties (1) and (2) ensure the existence of  a unique strong solution to the SDE (\ref{freedyn1_dim}) and 
the existence of a unique invariant probability measure, whereas  
property (3) ensures the ultracontractivity of the  associated semigroup, see \cite{KKR93}. 

\begin{EXA} (Markovian case)
Let $U$ satisfy above assumptions (1)-(3) and $b$ be a Markovian finite range bounded drift. It thus satisfies {\em (B2)} and {\em (B3)}. 
This case includes the one treated in {\em \cite{DR05}}. 
\end{EXA}
\begin{EXA} (Stochastic resonance) \label{resonance}
One can generalize the free dynamics in such a way that it remains Markovian but is no more time-homogeneous, introducing an
external periodic signal in the dynamics {\em (\ref{freedyn1_dim})}.
These models are used to describe the so-called {\em stochastic resonance} effect, see e.g. {\em \cite{WSB04, B10, KLYMY10}}. 
So, let us  consider as concrete example the following dynamics
\begin{equation}
dx(t) =   dB(t) - \frac{1}{2}\Big( x^3(t) - x(t)  - A \sin(t) \Big)dt , 
\end{equation}
where the drift derives from a time-independent  potential  given by 
$
U(x) := \frac{1}{4}x^4 - \frac{1}{2}x^2 
$
together with a bounded time-periodic forcing with amplitude $A>0$.
In that case properties {\em (1)-(3)} are satisfied.
\end{EXA}
\begin{EXA} (Free dynamics with delay)
One can generalize the free dynamics introducing a delayed feedback. It then becomes non Markovian. 
The over-damped particle motion in the double-well quartic potential as introduced 
 in {\em \cite{TP01}} furnishes such an  example:
\begin{equation}
dx(t) =  dB(t) - \frac{1}{2}\Big( x^3(t) - x(t)  - \alpha x(t-t_0)\Big)dt \, ,
\end{equation}
where $\alpha > 0$ is the strength of the feedback. 
\end{EXA}
The following examples are non-Markovian since they include a time  memory.
\begin{EXA} (Independent dynamics with time memory)
Let $U$ satisfy {\em (1)-(3)}. We define the drift by
\begin{equation}
b_i(t,\omega) := \begin{cases}  \int_0^{t} \epsilon(s) f(\omega_i(s)) ds & \text{ if } t < t_0  \\
\int_{t-t_0}^t \epsilon(s) f(\omega_i(s)) ds & \text{ if } t \geq t_0 
 \end{cases}
\end{equation}
where $f:\mathbb{R}\rightarrow \mathbb{R}$ is a measurable bounded function and the time-memory function 
$\epsilon:[0,\infty)\rightarrow \mathbb{R}$ is assumed
to be
integrable. This kind of drift $b$ is non-Markovian since it depends on a finite time window with length $t_0$.
\end{EXA}
\begin{EXA} (Interaction with finite extent in space and time)
Let $U$ satisfy {\em (1)-(3)}.  Fix $t_0 > 0$ and define the dift by
\begin{equation}
b_i(t,\omega) := 
\begin{cases} \int_0^{t} \alpha_i(t-s,\omega(s)) dV_s
& \text{ if } t < t_0  \\
\int_{t-t_0}^{t} \alpha_i(t-s,\omega(s)) dV_s
& \text{ if } t \geq t_0  
\end{cases}
\end{equation}
where the bounded variation integrator $V_s$ can be deterministic or stochastic and adapted. The functions $\alpha_i$ are bounded and spatially local:
\begin{equation}
\alpha_i(\cdot,x) = \alpha_i(\cdot,x_{\mathcal{N}}).
\end{equation}
Therefore $b$  depends on a finite time window with length $t_0$.
\end{EXA}
\subsubsection{Planar rotors}
In this section we would like to discuss how the above result for propagation of Gibbsianness can be adapted to planar rotors diffusions
with non-Markovian drift. It leads to a generalization of the conservation results presented in \cite{VER09}, where the authors considered Markovian dynamics.

Let us first introduce the setting. Take now  $\mathbb{S}^{\mathbb{Z}^d}$ as configuration space where $\mathbb{S}$ is the unit circle, which  we
can identify with the space interval $[0,2\pi)$ where $0$ and $2\pi$ are considered to be the same points. We consider the
solution $X^{\odot}=(X^{\odot}_i(t))_{t\geq 0,i\in \mathbb{Z}^d}$ of the following infinite system of Stochastic Differential Equations
\begin{equation}
\begin{cases}
& dX^{\odot}_i(t)  = dB^{\odot}_i(t) + \beta \  b_i(t,X^{\odot}) dt, i \in \mathbb{Z}^d \\
& X^{\odot}(0)  \sim \nu,  \label{SDEcircle}
\end{cases}
\end{equation}
on the path space $\Omega_S := C(\mathbb{R}_+,\mathbb{S})^{\mathbb{Z}^d}$ endowed  by the canonical sigma-field $\mathcal{F}$. $(B^{\odot}_i(t))_{t\geq 0, i
\in \mathbb{Z}^d}$ is a sequence of independent Brownian motions living on the circle $\mathbb{S}$ and the drift term of the $i^{th}$ coordinate, again denoted
by 
$b_i(t,\cdot)$, can depend on the values of the other coordinates on the whole time-interval $[0,t]$. 
Furthermore $\nu$ is supposed to be a suitable initial Gibbs measure. Let $Q^{\nu}$ denote the law of the solution of the SDE  \eqref{SDEcircle} with initial
measure $\nu$.\\

In the following let us present our assumptions. \\
The interaction defining the initial  Gibbs measure is supposed to be strong summable, that is it satisfies (A1). 
In the framework of planar rotors, since $\mathbb{S}$ is compact, the class of such interactions is indeed much larger than for unbounded spins. \\
The circle is the simplest compact manifold, hence we get immediately the ultracontractivity of the  semigroup associated to the free dynamics,
see for example \cite{GZ02} Theorem 3.3 and exercise 3.8.
\\
We assume that the space-time interactions $b_i$ are local in space and time and bounded, that is it satisfy  assumption (B2) and (B3). 

Then we can formulate our result in the context of planar rotors. Its proof follows the same steps as in section \ref{sec:laws}-\ref{sec:doublelayer}, hence we
will not repeat it here.

\begin{THM} \label{main}
Consider $Q^{\nu}$, the law of the infinite-dimensional SDE \eqref{SDEcircle} 
with a drift satisfying assumptions {\em (B2)} and {\em(B3)} and suppose that the initial distribution 
 $\nu$ is a Gibbs measure in $\mathcal{G}_{\beta_0 }(\phi)$ where $\phi$ satisfies the strong summability assumption {\em (A1)}.
There exists a bound $\bar \beta_0>0$ for the initial inverse temperature and a bound  $\bar \beta > 0$ 
for the intensity of the space-time interaction such
that,
if  $0 \leq \beta \leq \bar\beta$ and $ 0 \leq \beta_0
\leq \bar\beta_0$, for all $t \geq 0$ the time-evolved
measure $Q^{\nu}\circ X(t)^{-1}$ is a Gibbs measure  w.r.t. some interaction $\phi^t$, which is then absolutely summable.
\end{THM}

\begin{COR}
The above Theorem \ref{main} provides a constructive way to obtain a solution of the SDE \eqref{SDEcircle} at any time $t$ for small $\beta$ as limit (in terms
of cluster expansions) of finite-dimensional approximations, whose existence (and uniqueness) is ensured by the assumption  {\em (B3)}.
\end{COR}


\begin{thebibliography}{99}
\bibitem[AHMP07]{AHMP07}
M. Arriojas, Y. Hu, S.-E. Mohammed, S.-E., G. Pap,
\textit{A delayed Black and Scholes formula}, Stoch. Ana. and Appl. 25, p.471--492, 2007.

\bibitem[B10]{B10}
R. Benzi, \textit{Stochastic resonance: from climate to biology}, Nonlin. Proc. Geophys. 17, p.431--441, 2010.




\bibitem[DPR04]{DPR04}
P. Dai Pra, S. R\oe lly, \textit{An existence result for infinite-dimensional Brownian diffusions with non-regular and non-Markovian drift},
Mark. Proc. Rel. Fields 10, p.113--136, 2004.



\bibitem[DPRZ02]{DPRZ02}
P. Dai Pra, S. R\oe lly, H. Zessin 
\textit{A Gibbs variational principle in space-time for infinite dimensional diffusions},
Probab. Th. Rel.Fields, 122, p.289--315, 2002.

\bibitem[DR05]{DR05}
D. Dereudre, S. R\oe lly, \textit{Propagation of Gibbsianness for Infinite-dimensional Gradient Brownian Diffusions},
Jour. Stat. Phys. Vol 121, p.511--551, 2005.

\bibitem[DvGVW95]{DvGVW95}
O. Diekmann, S.A. Van Gils, S.M. Verduyn Lunel, H.O. Walther,
\textit{ Delay Equations}, Functional-, Complex-,
and Nonlinear Analysis, Springer-Verlag, New York, 1995.

\bibitem[EK10]{EK10}
V. Ermolaev, C. K\"ulske, \textit{Low-temperature dynamics of the Curie–Weiss model: periodic orbits, multiple histories, and loss of Gibbsianness}, 
J. Stat. Phys. 141, p.727--756, 2010.


\bibitem[G88]{G88}
H.-O. Georgii, \textit{Gibbs measures and phase transitions}, Berlin, W. de Gruyter, 1988.


\bibitem[GZ02]{GZ02}
A. Guionnet, B. Zegarlisnki, \textit{Lectures on
Logarithmic Sobolev Inequalities}, Sem. de prob. H. Poincare, tome 36, p.1--134, 2002.

\bibitem[HV05]{HV05}
V. Blaka Hallulli, T. Vargiolu, 
\textit{Financial models with dependence on the past: a survey}, 
In Applied and Industrial Mathematics in Italy 348 - 359,
M. Primicerio, R. Spigler, V. Valente, editors, Series on Advances in Mathematics for Applied Sciences Vol. 69, World Scientific 2005


\bibitem[KLYMY10]{KLYMY10}
Y. Kai-Leung, L. You-Ming, X. Yan, \textit{Stochastic resonance in the FitzHugh–Nagumo system driven by bounded noise}, Chin. Phys. B 19, 010503, 2010.


\bibitem[KKR93]{KKR93}
O.Kavian, G. Kerkyacharian, B. Roynette, \textit{Quelques remarques sur l'ultracontractivit\'e}, J. Func. Anal. 111, p.155--196, 1993.

\bibitem[KLN07]{KLN07}
C. K\"ulske, A. Le Ny, \textit{Spin-flip dynamics of the Curie–Weiss model: loss of Gibbsianness with possibly broken symmetry}, Comm. Math. Phys. 271, no 2 ,
p.431--454, 2007.

\bibitem[KP86]{KP86}
R. Koteck\'y, D. Preiss, \textit{cluster expansions for abstract polymer models}, Comm. Math. Phys. 103, p.491--498, 1986.

\bibitem[K74]{K74}
O. K. Kozlov, \textit{Gibbs description of a system of random variables}, Prob. Info. Trans. 10, p.258--265, 1974.

\bibitem[KR06]{KR06}
C. K\"ulske, F. Redig, \textit{Loss without recovery of Gibbsianness
during diffusion of continuous spins}, Probab. Theory Relat. Fields 135, p.428--456, 2006.

\bibitem[MVZ00]{MVZ00}
R.A. Minlos, A. Verbeure, V. Zagrebnov, \textit{A quantum crystal model in the light mass limit: Gibbs states}, Rev. Math. Phys., Vol. 12-7, p.981--1032, 2000.

\bibitem[RRR10]{RRR10}
F. Redig, S. Roelly, W. Ruszel, \textit{Short-time Gibbsianness for infinite-dimensional diffusions with space-time interaction}, J. Stat. Phys. 138, no 6,
p.1124--1144, 2010.

\bibitem[TP01]{TP01}
L.S. Tsimring, A. Pikovsky, \textit{Noise-induced dynamics in bistable systems with delay}, Phys. Rev. Lett. 87, 250602, 2001.

\bibitem[VEEIK12]{VEEIK12}
A.C. D. van Enter, V. N. Ermolaev, G. Iacobelli, C. K\"ulske,
\textit{Gibbs–non-Gibbs properties for evolving Ising models on trees}, Ann. de l'inst. Henri Poinc. Prob. et Stat. Vol 48, no 2, p.774--791, 2012.

\bibitem[VEFHR02]{VEFHR02}
A.C.D. van Enter, R. Fern\'andez, F. den Hollander, F. Redig,
\textit{Possible loss and recovery of Gibbsianness during the stochastic
evolution of Gibbs measures}, Comm. Math. Phys. 226, no 1,
p.101--130, 2002.

\bibitem[VEFHR10]{VEFHR10}
A.C.D. van Enter, R. Fern\'andez, F. den Hollander, F. Redig,
\textit{A large-deviation view on dynamical Gibbs–non-Gibbs transitions},  
Mosc. Math. J. Vol.10, no 4, p.687--711, 2010. 

\bibitem[VEFS93]{VEFS93}
A.C.D. van Enter, R. Fern\'andez, A.D. Sokal, \textit{Regularity properties and pathologies of position-space renormalization-group
transformations: Scope and limitations of Gibbsian theory}, J.
Stat. Phys. 72, p.879--1167, 1993.


\bibitem[VER08]{VER08}
 A.C.D. van Enter, W.M. Ruszel,
\textit{Loss and recovery of gibbsianness for xy spins in a small external field},      J. Math. Phys. 49,125208-1/8, 2008. 

\bibitem[VER09]{VER09}
 A.C.D. van Enter, W.M. Ruszel,
\textit{Gibbsianness versus Non-Gibbsianness of time-evolved planar rotor models}, Stoch. Proc. Appl. 119 , p.1866--1888, 2009.
      

\bibitem[WSB04]{WSB04}
T. Wellens, V. Shatokhin, A. Buchleitner, \textit{Stochastic resonance}, Rep. Prog. Phys. 67, 45, 2004.
\end{thebibliography}
\end{document}